\documentclass{llncs}
\usepackage{epsf}
\usepackage{graphicx}
\usepackage{url}

\newtheorem{prop}{Proposition}
\newtheorem{df}{Definition}

\title{A Logic Programming Framework for Combinational Circuit Synthesis}

\author{
   {\bf Paul Tarau}\\
   Department of Computer Science and Engineering\\
   University of North Texas\\
   {\em paul.tarau@gmail.com}\\
~~\\   
   {\bf Brenda Luderman}\\
   Intel Corp.\\
   Austin, Texas\\
   {\em brenda.luderman@gmail.com}\\
}

\institute{}

\date{} 

\begin{document}
\thispagestyle{empty}
\maketitle

\begin{abstract}
Logic Programming languages and combinational circuit synthesis tools share a common ``combinatorial search over logic formulae" background.
This paper attempts to reconnect the two fields with a fresh look at Prolog encodings for the combinatorial objects involved in circuit synthesis. While benefiting from Prolog's fast unification algorithm and built-in backtracking mechanism, efficiency of our search algorithm is ensured by using parallel bitstring operations together with logic variable equality propagation, as a mapping mechanism from primary inputs to the leaves of candidate Leaf-DAGs implementing a combinational circuit specification. After an exhaustive expressiveness comparison of various minimal libraries, a surprising first-runner, Strict Boolean Inequality ``$<$" together with constant function ``$1$" also turns out to have small transistor-count implementations, competitive to NAND-only or NOR-only libraries. As a practical outcome, a more realistic circuit synthesizer is implemented that combines rewriting-based simplification of $(<,1)$ circuits with exhaustive Leaf-DAG circuit search. 

\end{abstract}

\section{Introduction}
Various logic programming applications and circuit synthesis tools share algorithmic techniques ranging from search over combinatorial objects and constraint solving to symbolic rewriting and code transformations.

The significant semantic distance between the two fields, coming partly from the application focus and partly from the hardware/software design gap has been also widened by the use of lower level procedural languages for implementing circuit design tools - arguably for providing better performance fine tuning opportunities. 

While intrigued by the semantic and terminological gap between the two fields, our interest in the use of logic programming for circuit design has been encouraged because of the following facts:

\begin{itemize}
\item the simplicity and elegance of combinatorial generation algorithms in the context of Prolog's backtracking, unification and logic grammar mechanisms
\item the structural similarity between Prolog terms and the Leaf-DAGs typically used as a data structure for synthesized circuits
\item elegant implementations of circuit design tools in high level functional languages \cite{odonnell}
\item the presence of new flexible constraint solving Prolog extensions like {\em CHR} \cite{chr94} that could express layout, routing and technology mapping aspects of the circuit design process needed, besides circuit synthesis, for realistic design tools.
\end{itemize}

The paper summarizes our efforts on solving some realistic combinational circuit synthesis problems with logic programming tools. It is organized as follows.
Section \ref{comb} describes the generation of combinatorial objects needed for exact circuit synthesis in Prolog, section \ref{bits} shows uniform bitstring representations for functions and primary inputs that check function equivalence without backtracking. 
Section \ref{libs} compares various universal boolean function libraries in terms of total cost of minimal representations of the set of 16 2-argument operators as an indicator of expressiveness for minimal cost synthesis purposes.
As result of this comparison, section \ref{strict} focuses on a surprisingly interesting library consisting of Strict Boolean Inequality and constant function $1$ with subsection \ref{univ} showing universality of $(<,1)$ and subsection \ref{rew} presenting our library specific rewriting algorithm, usable as minimization heuristics when exact synthesis becomes intractable.
Section \ref{tran} describes low transistor-count implementations of the $<$-gate and compares transistor counts for $(<,1)$ with equivalent NAND-based circuits. Sections \ref{rel} and \ref{fwork} discuss related and future work and section \ref {conc} concludes the paper. The Prolog code for the exact synthetizer and various libraries is available at \url{http://logic.csci.unt.edu/tarau/research/2007/csyn.zip}.

\section{Combinatorial Objects and Combinational Circuit Synthesis} \label{comb}

Our exact synthesis algorithm uses Prolog's depth-first backtracking to find minimal circuits representing boolean functions, based on a given library of primitives, through composition of combinatorial generation steps and efficient checking against an output pattern specified as a truth table.

The general structure of the algorithm is as follows:

Through the paper, Leaf-DAGs will be used to represent the synthetized circuits. The general structure of the algorithm is as follows:

\begin{enumerate}
\item First, the algorithm runs a library specific rewriting module (see \ref{rew} for a library specific rewriting module) on the input formula (in symbolic, CNF or DNF form). This (or a conservative higher estimate) provides an upper limit (in terms of a cost function, for instance the number of gates) on the size of the synthesized expression. It also provides a (heuristically) minimized formula, in therms of the library, that can be returned as output if exact synthesis times out.
\item Next, if the formula qualifies for exact synthesis, we enumerate candidate trees {\em in increasing cost order}, to ensure that minimal trees are generated first. This involves the following steps:

\begin{enumerate}
\item First, we implement a mapping from the set primary inputs to the set of their occurrences in a tree. This involves generating functions from N variables to M occurrences. We achieve this without term construction, through logical variable bindings, efficiently undone on backtracking by Prolog's trailing mechanism.
\item Next, N-node binary trees of increasing sizes are generated. The combination of the expression trees and the mapping of logic variables (representing primary inputs) to their (possibly multiple) occurrences, generates Leaf-DAGs.
\end{enumerate}

\item Finally, we evaluate candidate Leaf-DAGs for equivalence with the output specification.
\end{enumerate}

We will describe the details of the algorithm and the key steps of their Prolog implementation in the following subsections.

\subsection{Boolean Expression Trees} \label{trees}

Size-constrained expression trees are combinatorial objects providing the skeletons for the Leaf-DAGs generated by our algorithm, as shown in predicate \verb~enumerate_tree_candidates/5~. The constraints are expressed as input parameters {\tt UniqueVarAndConstCount} and {\tt LeafCount}. The generator produces an expression tree {\tt ETree} and computes its truth table {\tt OutSpec} encoded as a bitstring-integer. Size-constraints, ensuring termination of the recursive predicate \verb~generate_expression_tree/7~, are encoded as a finite list of nodes, using DCG notation. Termination is ensured by having each recursive step consume exactly one node.  A finite list of leaf variables provides leaves to the generated tree in the first clause of predicate \verb~generate_expression_tree/7~.

{\small \begin{verbatim}
enumerate_tree_candidates(UniqueVarAndConstCount,LeafCount, 
                                          Leaves,ETree,OutputSpec):-
  N is LeafCount-1,
  length(Nodes,N),
  generate_expression_tree(UniqueVarAndConstCount,ETree,OutputSpec,
                                                  Leaves,[],Nodes,[]).

generate_expression_tree(_,V,V,[V|Leaves],Leaves)-->[].
generate_expression_tree(NbOfBits,ETree,OutputSpec,Vs1,VsN)-->[_],
  generate_expression_tree(NbOfBits,L,O1,Vs1,Vs2),
  generate_expression_tree(NbOfBits,R,O2,Vs2,VsN),
  {combine_expression_values(NbOfBits,L,R,O1,O2,ETree,OutputSpec)}.
\end{verbatim}}

\noindent The predicate \verb~combine_expression_values/7~, shown below for the $(*,\oplus,1)$ library, produces tree nodes like \verb~L*R~ and \verb~L^R~, while computing their bitstring-integer encoded truth table {\tt O} from the left and right branch values {\tt O1} and {\tt O2}.

{\small \begin{verbatim}
combine_expression_values(_,L,R,O1,O2, L*R,O):-bitand(O1,O2,O).
combine_expression_values(_,L,R,O1,O2, L^R,O):-bitxor(O1,O2,O).
\end{verbatim}}

\noindent The generated trees have binary operators as internal nodes and variables as leaves. They are counted by Catalan numbers \cite{EIScatalan}), with $4^N$ as a (rough) upper bound for N leave trees.

\subsection{Implementing Finite Functions as Logical Variable Bindings} \label{fun}

We express finite functions as {\em bindings} of a list of logic variables (the range of the function) to values in the domain of the function.

{\small \begin{verbatim}
functions_from([],_).
functions_from([V|Vs],Us):-member(V,Us),functions_from(Vs,Us).

Example: A call like

?- functions_from([A,B,C],[0,1])

enumerates the 8 functions as variable bindings like:

{A->0,B->0,C->0}
{A->0,B->0,C->1}
...
{A->1,B->1,C->1}
\end{verbatim}}

\noindent Assuming the first set has M elements and the second has N elements, a total of $N^M$ backtracking steps are involved in the enumeration, one for each function between the two sets. As a result, a finite function can be seen simply as a set of variable occurrences. This provides fast combinatorial enumeration of function objects, for which backtracking only involves trailing of variable addresses and no term construction.

\subsection{Leaf-DAG Circuit Representations for Combinational Logic} \label{leaf}

\begin{df}
A {\em Leaf-DAG} is a directed acyclic graph where only vertices (called leaves) that have no outgoing edges can have multiple incoming edges. 
\end{df}

Leaf-DAGs can be seen as trees with possibly merged leaves. Note that Leaf-DAGs are naturally represented as Prolog terms with multiple occurrences of some variables - like $X$ and $Y$ in $f(X,g(X,Y,Z),Y)$.

Our Leaf-DAG generator combines the size-constrained tree generator from subsection \ref{trees} and the functions-as-bindings generator from subsection \ref{fun}, as follows:

{\small \begin{verbatim}
generate_leaf_dag(UniqueVarAndConstCount,LeafCount,
                  UniqueVarsAndConsts,ETree,OutputSpec):-
  length(Leaves,LeafCount),
  functions_from(Leaves,UniqueVarsAndConsts),
  enumerate_tree_candidates(UniqueVarAndConstCount,LeafCount,
                                      Leaves,ETree,OutputSpec).
\end{verbatim}}

\begin{prop}  \label{steps}
Let $catalan(M)$ denote the M-th Catalan number. The total backtracking steps for generating all Leaf DAGs with N primary inputs and M binary operation nodes is $catalan(M)*N^{M+1}$. 
\end{prop}

\begin{proof}
It follows from the fact that Catalan numbers count the trees and $N^{M+1}$ counts the functions corresponding to mapping the primary inputs to their leaves, because a binary tree with $M$ internal nodes, each corresponding to an operation, has $M+1$ leaves.
\end{proof}

Note that if constant functions like 0 or 1 are part of the library, they are simply added to the list of primary inputs.

The predicate \verb~synthesize_leaf_dag/4~ describes a (simplified version) of our Leaf-DAG generator. Note that if the OutputSpec truth table is given as a constant value, unification ensures that only LeafDAGs matching the specification are generated. With OutputSpec used as a free variable, the predicate can be used in combination with Prolog's dynamic database as part of a dynamic programming algorithm that tables and reuses subcircuits to avoid recomputation.

{\small \begin{verbatim}
synthesize_leaf_dag(MaxGates,UniqueBitstringIntVars,
                    UniqueVarAndConstCount,PIs:LeafDag=OutputSpec):-
  constant_functions(UniqueVarAndConstCount,ICs,OCs),
  once(append(ICs,UniqueBitstringIntVars,UniqueVarsAndConsts)),
  for(NbOfGates,1,MaxGates),
    generate_leaf_dag(UniqueVarAndConstCount,NbOfGates, 
                      UniqueVarsAndConsts,ETree,OutputSpec),
  decode_leaf_dag(ETree,UniqueVarsAndConsts,LeafDag,DecodedVs),
  once(append(OCs,PIs,DecodedVs)).
\end{verbatim}}

\begin{prop}
The predicate \verb~synthesize_leaf_dag/4~ generates Leaf-DAGs in increasing size order.
\end{prop}

\begin{proof}
It follows from the fact that each call to \verb~generate_leaf_dag/5~ enumerates on backtracking all Leaf-DAGs of size \verb~NbOfGates~ and the predicate \verb~for/3~ provides increasing  \verb~NbOfGates~ bounds.
\end{proof}

Assuming zero cost for constant functions and a fixed transistor cost for each operator, it follows that the synthesizer produces circuits based on {\em single-operator} libraries in increasing cost order. For more complex cost models adaptations to the tree generator can be implemented easily.
  
\section{Fast Boolean Evaluation with Bitstring Truth Table Encodings} \label{bits}

We use an adaptation of the clever bitstring-integer encoding described in the Boolean Evaluation section of \cite{knuth06draft} of $n$ variables as truth tables. Let $x_k$ be a variable for $0 \leq k<n$. Then $x_k={(2^{2^n}-1)}/{(2^{2^{n-k-1}}+1)}$, where the number of distinct variables in a boolean expression gives $n$, the number of bits for the encoding. The mapping from variables, denoted as integers, to their truth table equivalents, is provided by the following Prolog code:

{\small \begin{verbatim}
% Maps variable K in 0..Mask-1 to truth table 
% Xk packed as a bitstring-integer.
var_to_bitstring_int(NbOfBits,K,Xk):-
  all_ones_mask(NbOfBits,Mask),
  NK is NbOfBits-(K+1),
  D is (1<<(1<<NK))+1,
  Xk is Mask//D.
  
% Mask is a bitstring-integer ending with NbOfBits of the form 
% 11...1. It also provides an encoding of constant function 1.  
all_ones_mask(NbOfBits,Mask):-Mask is (1<<(1<<NbOfBits))-1. 
\end{verbatim}}

\noindent Variables representing such bitstring-truth tables can be combined with the usual bitwise integer operators to obtain new bitstring truth tables encoding all possible value combinations of their arguments, like in:
 
{\small \begin{verbatim}
bitand(X1,X2,X3):-X3 is '/\'(X1,X2).
bitor(X1,X2,X3):-X3 is '\/'(X1,X2).
bitxor(X1,X2,X3):-X3 is '#'(X1,X2).
bitless(X1,X2,X3):-X3 is '#'(X1,'\/'(X1,X2)).
bitgt(X1,X2,X3):-X3 is '#'(X1,'/\'(X1,X2)).
bitnot(NbOfBits,X1,X3):-all_ones_mask(NbOfBits,M),X3 is '#'(X1,M).
biteq(NbOfBits,X,Y,Z):-all_ones_mask(NbOfBits,M),Z is '#'(M,'#'(X,Y)).
bitimpl(NbOfBits,X1,X2,X3):-bitnot(NbOfBits,X1,NX1),bitor(NX1,X2,X3).
bitnand(NbOfBits,X1,X2,X3):-bitand(X1,X2,NX3),bitnot(NbOfBits,NX3,X3).
bitnor(NbOfBits,X1,X2,X3):-bitor(X1,X2,NX3),bitnot(NbOfBits,NX3,X3).
\end{verbatim}} 
 
\noindent The length of the bitstring-truth tables is sufficient for most perfect synthesis problems involving exhaustive search, as most problems become intractable above the 64 bits corresponding to 6 variables (see Proposition \ref{steps}). However, using arbitrary length integer packages, available for most Prologs, allows extending the mechanism further. In practice, a timeout mechanism can be used to decide if falling back to a heuristic synthesis method is needed.

\section{A Comparison of Universal Boolean Function Libraries} \label{libs}

\begin{df}
A set of boolean functions $F$ is universal if any boolean function can be written as a composition of functions in $F$.
\end{df}

A well known universal set is (conjunction, negation) i.e. $(*,\sim)$ - this follows immediately from the rewriting of a truth table in terms of conjunction, disjunction and negation followed by elimination of disjunctions using De Morgan's laws. Universality of a library is usually proven by expressing conjunction and negation with its operations.

The table in Fig. \ref{exp} compares a few libraries used in synthesis with respect to the total gates needed to express all the 16 2-argument boolean operations (themselves included). The last column marks if the library is {\em non-redundant} (or {\em minimal}), i.e. if none of its functions can be expressed in terms of the others.

\begin{figure}[htbp]
\centering
~~\\
\begin{tabular}{|c|c|c|}
\hline
Library & total for 16 operators & non-redundant\\
\hline
$nand$ & 46 & yes\\
$nor$ & 46 & yes \\
$nand,1$ & 33 & no\\
$nor,0$ & 33 & no \\
$*,nand$ &32 & no\\
$<,nor$ & 31 & no\\
$\Rightarrow,0$ & 28 & yes\\
${\bf <,1}$ & {\bf 28} & {\bf yes}\\
$*,<,1$ & 26 & no\\
$*,\oplus,1$&25 & yes\\
$<,nand,1$ & 25 & no\\
$<,nor,1$&24 &no\\
$*,=,0$ & 23 &yes\\
$\Rightarrow,=,0$ & 21 &no\\
$<,=,1$ & 21 &no\\
\hline
\end{tabular}
~~\\
\caption{Relative Expressiveness of Libraries}
\label{exp}
\end{figure}

The comparison gives a hint on the relative expressiveness of libraries. 

By including operations like ``$\oplus$" and ``$=$", that are known to require a relatively high number of other gates (or a high transistor count) to express, one can minimize the number of operators (and circuit size) required. Using only gates known to have low transistor-count implementations like {\tt nand} and {\tt nor}, the expressiveness drops significantly (46 required). 

Surprisingly, $(\Rightarrow,0)$ and its dual $(<,1)$ do significantly better than {\tt nand} and {\tt nor}: they can express all 16 operators with only 28 gates. As section \ref{tran} will show, $(<,1)$ turns out to have very interesting low transistor implementations. Given also that it has not been used in any work on synthesis that we are aware of, we will explore this library in depth in section \ref{strict}. 

Interestingly enough, the libraries $(*,=,0)$ and $(\Rightarrow,=,0)$ that provide, arguably, some the most human readable expressions when expressing other operators,  have relatively small gate counts (23 and 21). For instance the first one expresses $A\Rightarrow B$ as $A=A*B$ and $A \oplus B$ as $(A=B)=0$, the second one expresses $(A+B)$ as $(0=A)=>B$, and both express $(A \oplus B \oplus C)$ as $(A=(B=C))$.

Note also, that besides spotting out the minimal universal library $(<,0)$, the comparison also identifies $(<,nor,1)$ as a highly expressive library, with potential for practical design uses, given that $<$ and $nor$ have both low transistor-count implementations.

Finally, one of the overall ``winners" of the comparison is $(<,=,1)$. It expresses the 16 operators with only 21 gates and it is a superset of the $(<,1)$ library. This also suggests exploring in more detail the potential of $<$ for synthesis.

\section{Using Strict Boolean Inequality for Combinational Circuit Synthesis} \label{strict}

While Strict Boolean Inequality\footnote{(Equivalent to ($\sim A)*B$ as well as $ \sim (A \Leftarrow B)$). Called Converse Nonimplication as well as ''NOT A BUT B" by Knuth \cite{knuth06draft}. Also called NIF standing for NOT (A IF B) and Half-XOR.} $A<B$ together with $1$ is a universal boolean function pair, it has been neglected by logicians as well as circuit designers, to the point where there are surprisingly few references to it in the literature in both fields.
Interestingly enough, its {\em dual}, $(\Rightarrow,0)$\footnote{Logical Implication with Falsehood (also denoted $\bot$)} -- is a well known universal function pair that has been extensively studied as an axiomatic basis for both classical and intuitionistic propositional logic.

One can only speculate about the reasons for this neglect. The lack of algebraic grouping properties like commutativity and associativity comes to mind. Or, that its intuitive meaning would be harder to map to common reasoning patterns.

In any case, none of these are critical for the synthesis problem, which, stated generically, is about {\em finding minimal representations of finite functions}\footnote{All finite functions can be expressed as boolean functions, by using binary encodings of their arguments and values.} in terms of a {\em universal subset} of them, given as a {\em library}.

As an indication of the usefulness of $(<,1)$ for synthesis, let's note that $A \oplus B$ (known to be part of notoriously hard to synthesize boolean functions) is in fact $(A<B)+(B<A)$ and therefore $A<B$ can provide half of $A \oplus B$. Note that it also provides a form of conjunction (with first argument inverted), given its equivalence to $\sim~A*B$. It follows from this equivalence, that $<$ also works as an inference rule: from its truth, one can determine uniquely the truth values of both of its arguments, i.e. the first should be {\tt false} and the second {\tt true}. As a side note, the reader might notice that this is similar, but in a way stronger than the mechanism through which {\em Modus Ponens} works. In the case of Modus Ponens, if one looks at its premises as a formula, then $A*(A \Rightarrow B)$ is equivalent to $A*B$ implying the truth of $B$ in addition to the (already assumed) truth of $A$. The key difference, that makes Modus Ponens more intuitive is, of course, that it provides an inference mechanism that conserves and extends truth, while using $A<B$ as an inference mechanism would force one to deal with both true and false consequences. 

\subsection{Strict Boolean Inequality as a Universal Boolean Operator}\label{univ}

\begin{df}
Strict Boolean Inequality is defined by the following truth table:
~~\\
\begin{tabular}{|c|c|c|}
\hline
$A$ & $B$ & $A<B$\\
\hline
0 & 0 & 0\\
0 & 1 & 1\\
1 & 0 & 0\\
1 & 1 & 0\\
\hline
\end{tabular} 
\end{df}

\begin{prop}
Strict Boolean Inequality $A<B$ together with constant function $1$ is a universal boolean function. 
\end{prop}
\begin{proof}
Given that conjunction and negation form a universal boolean function pair, the proposition follows from the fact that conjunction $A*B$ has the same truth table as $(A < 1)<B$ and that negation $\sim A$ has the same truth table as $(A<1)$.
\end{proof}

\subsection{The Symbolic Rewriting Algorithm}\label{rew}

Our symbolic rewriting recurses over a given formula, and after each rewriting step, it proceeds with simplifications using propositional tautologies. We will illustrate the algorithm with the table in Fig. \ref{rr} showing how various expressions are transformed after the recursive rewriting of their arguments. For a given argument $A$ we denote $`A$ the result of recursive application of the algorithm to $A$. The algorithm preserves constants and primary input variables unchanged. We also assume that simplification occurs {\em implicitly} after each transformation step.

\begin{figure}[htbp]
\centering
~~\\
\begin{tabular}{|c|c|}
\hline
From & To\\
\hline
$0$ & $0$\\
$1$ & $1$\\
$A<B$ & $`A< `B$\\
$\sim A$ & $`A<1$\\
$A \Leftarrow B$ & $(`A< `B)<1$\\
$A*B$ & $(`A<1)<`B$\\
$nor(A,B)$ & $`A<(`B<1)$\\
$A+B$ & $`(A \Leftarrow (\sim B))$\\
$A \Rightarrow B$ & $ (`B < `A)<1$\\
$A \oplus B$ & $(`A<`B)+(`B<`A)$\\
$A=B$ & $`(nor((A<B),(B<A))<1)$\\
$ite(C,T,F)$ & $`((C \Rightarrow T)*(\sim C \Rightarrow F))$\\
\hline
\end{tabular}
~~\\
\caption{$(<,1)$-Rewriting Rules}
\label{rr}
\end{figure}

The algorithm reduces most simple tautologies to $1$ and most simple contradictions to $0$. As a result, it also may reduce the number of variables on which the expression actually depends. 

\paragraph*{Optimizing for Minimal Transistor Count} Given that constant function $1$ is 0-cost and that function $<$ has a 4-transistor cost (see section \ref{tran}), the synthesis algorithm can assume that the cost is given by the number of $<$ gates.

\paragraph*{Delay-Constrained Minimal Circuit Synthesis} Given the uniform gate structure of the circuits, we can ensure that delays are within acceptable margins by simply constraining the maximum length of the longest path from the primary inputs to the primary outputs.

\subsection{Minimal $(<,1)$-representations for Key Boolean Functions} \label{min}

Figure \ref{minrep} shows minimal representations for 0, negation, some 2-input boolean functions and the 3-argument IF-THEN-ELSE, as produced by our synthesizer. Interestingly enough, the minimal formulae obtained by exhaustive search are identical (as in the case of most simple formulae) with those obtained using our rewriting algorithm.

\begin{figure}[htbp]
\centering
~~\\
\begin{tabular}{|c|c|}
\hline
$Function$ & $``<" Representation$ \\
\hline
$0$ & $1<1$\\
$\sim A$ & $A<1$\\
$A*B$ & $(A < 1) < B$ \\
$A+B$ & $ (A < (B < 1)) < 1$ \\
$A \Rightarrow B$ & $(B < A) < 1$ \\
$A \Leftarrow B$ & $(A < B) < 1$ \\
$A \oplus B$ & $((A < B) < ((B < A) < 1)) < 1$ \\
$A=B$ & $(A < B) < ((B < A) < 1)$ \\
$A$ NAND $B$ & $((A < 1) < B) < 1$ \\
$A$ NOR $B$ & $A < (B < 1)$ \\
IF $A$ THEN $B$ ELSE $C$ & $(A < (C < 1)) < ((B < A) < 1)$ \\
\hline
\end{tabular} 
\caption{$(<,1)$-Representations of some functions}
\label{minrep}
\end{figure}

\subsection{Synthesis from CNF and DNF forms} \label{cnf}

As Disjunctive Normal Forms (DNF, also called sum-of-products) and Conjunctive Normal Forms (CNF , also called product-of-sums) are the result of repeated conjunctions and disjunctions, we first focus on optimal $(<,1)$-representation of these.

\begin{prop}
A sequence of disjunctions of $N$ variables has a minimal $(<,1)$-representation with 2 occurrences of constant 1 and exactly one occurrence of each input variable, provided by the formula:

$A_1+A_2+ \dots +A_N=(A_1 < (A_2 < \dots (A_N < 1) \dots )) < 1$
\end{prop}

\begin{prop}
A sequence of conjunctions of $N$ variables has a minimal $(<,1)$-representation with $N-1$ occurrences of constant $1$ and exactly one occurrence of each input variable, provided by the formula:

$A_1*A_2* \dots A_{N-1}*A_N=((A_1 < 1) < ((A_2 < 1) < \dots ((A_{N-1} < 1) < A_N) \dots )$
\end{prop}
\begin{proof}
By induction on the number of input variables, N.
\end{proof}

Synthesis from CNF and DNF formulae (that can be obtained directly from truth table descriptions of circuits) proceeds by applying the encodings provided by the previous propositions recursively, followed by (and interleaved with) simplification steps.

\section{Transistor Implementations for $(<,1)$-circuits}\label{tran}

\begin{figure}[htbp]
\centering
\scalebox{0.30}{\includegraphics*[bb=0pt 0pt 600pt 600pt]{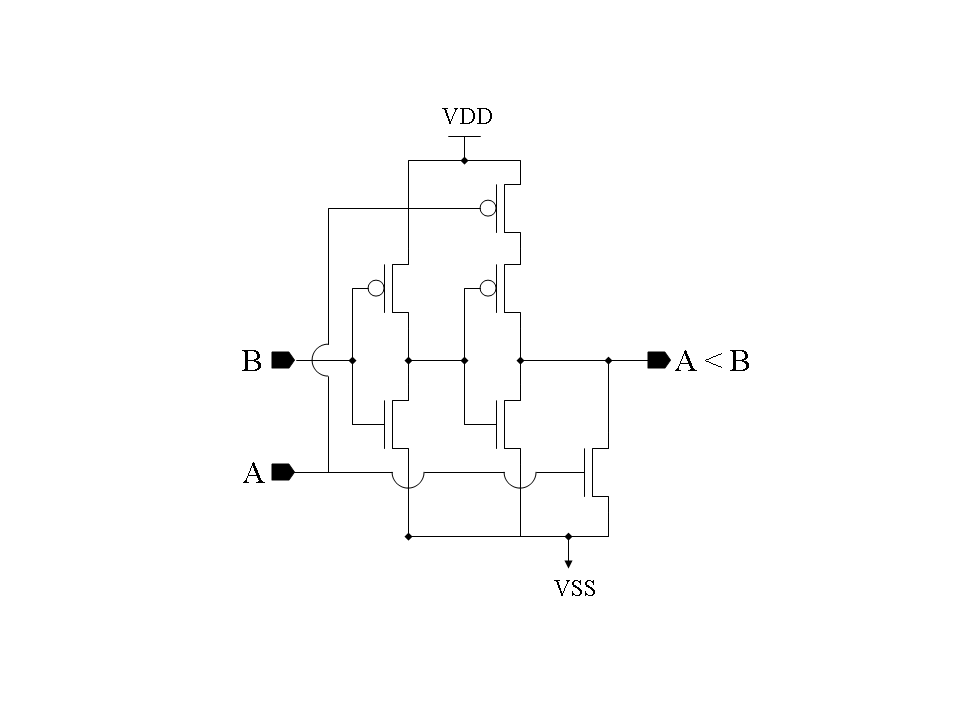}}
\caption{Powered 6-Transistor $A<B$}
\label{t6}
\end{figure}

Clearly as $A<B$ is equivalent to $nor(A,\sim B)$, an obvious 6-transistor implementation is obtained when input B drives a 2-transistor inverter while its output and input A drive a 4-transistor NOR gate. This logic circuit is shown in Fig. \ref{t6}. The output node, $A<B$, has a direct path to the power nodes VDD and VSS through the source connections of the transistors connected to it. As a result, the output is called ``buffered" and the logic circuit type is ``powered".
 
\begin{figure}[htbp]
\centering
\scalebox{0.30}{\includegraphics*[bb=0pt 0pt 600pt 600pt]{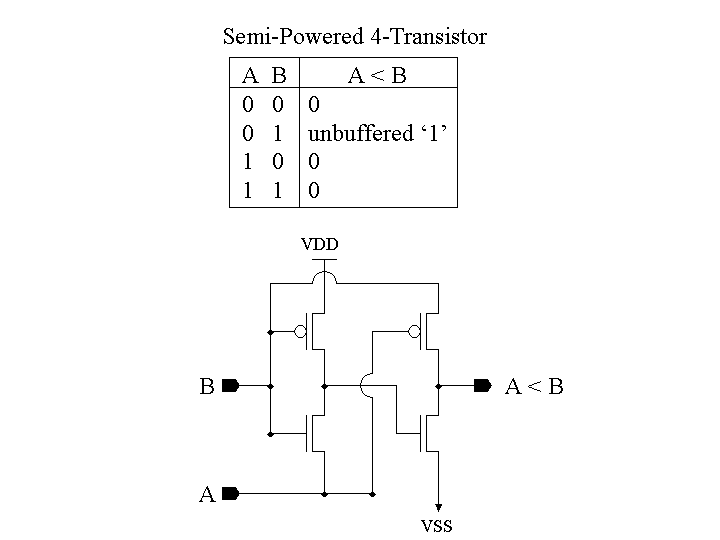}}
\caption{Semi-Powered 4-Trans. $A<B$}
\label{t4s}
\end{figure}

To reduce transistor count, a {\em pass transistor logic} (PTL) circuit for $A<B$ can be implemented using 4 transistors. In this circuit, the output node, $A<B$, in Fig. \ref{t4s} has a direct path to the power net VSS while input B provides the VDD power.  Therefore, the logic circuit type is ``semi-powered" and the output level for VDD is called ``unbuffered".

The constant function $1$ can be implemented by direct routing to the VDD power grid.
Similary, the constant function $0$ can be implemented by direct routing to the VSS power grid.

\noindent {\em In conclusion, assuming a design using PTL-logic, the transistor count for an implementation of the $<$ function is 4, while constant functions $1$ and $0$ are essentially free, with transistor count 0.}

\subsection*{Transistor Count Comparisons for $(<,1)$ and NAND.}

\begin{figure}[htbp]
\centering
~~\\
\begin{tabular}{|c|c|c|}
\hline
Function & $<$-cost & NAND-cost\\
\hline
$A=B$ & 4*4=16 & 5*4=20\\
$A\oplus B$ & 5*4=20 & 5*4=20\\
$A*B$ &       2*4= 8 & 3*4=12 (8 if sharing)\\ 
$(A*B) \Rightarrow C$ & 4*4=16 & 4*4=16\\
$A*B*C$ & 4*4=16 & 6*4=24\\ 
$A+B+C$ & 4*4=16 & 7*4=28\\ 
if-then-else & 5*4=20 & 4*4=16\\
$(A \Rightarrow B)*(B \Rightarrow C))$ & 4*4=16 & 5*4=20\\
$nand(A,B)$ & 3*4=12 & 1*4=4\\
$A<B$ & 1*4 & 5*4=20 (16 if sharing) \\
2x2 half-adder & 20+8=28 & 20+12=32\\
\hline
\end{tabular} 
~~\\
\caption{Transistor Costs}
\label{tc}
\end{figure}

To have a glimpse at the competitiveness of $(<,1)$ as a universal pair, in comparison with a minimal NAND-based circuit, we have compared costs obtained as transistor counts of the resulting circuits.

\noindent While the comparison only involves Leaf-DAG representations and ignores the fact that stronger sharing could be present in the case of multiple outputs or the case arbitrary DAGs are used, it shows, at a small scale, that practical uses of the $(<,1)$ for exact synthesis are likely to be competitive. Note also that in practice commutativity of NAND brings more sharing opportunities if general DAGs are used and that both $A<1$ and $nand(A,A)$ can be replaced with 2-transistor inverters.

\section{Related Work} \label{rel}

Mentions of Prolog for circuit simulation go back as early as \cite{clocksin:programming:87}. Peter Reintjes in \cite{reintjes92} mentions CMOS circuit design and Prolog as two {\em Elegant Technologies} with potential for interaction.
Knuth in \cite{knuth06draft}, section 7.1.2 mentions $A<B$ as forming one of the 5 (out of 16) boolean function used as part of a {\em boolean chain} (sequence of connected 2-argument boolean functions) needed for synthesis by exhaustive enumeration. Interestingly, the other 4 are: $>,*,+,\oplus$. Note that $>$ is the symmetric of $<$, and that with its exception, $*,+,\oplus$ have been heavily used in various synthesis algorithms. This supports our intuition that $<$ and $>$'s potential for synthesis is worth further exploration. Knuth also computes minimal representations of all 5-argument functions using a clever reduction to equivalence classes and proves that the cost for representing almost every boolean function of N-arguments in terms of boolean chains exceeds $2^N/N$.

Rewriting/simplification has been used in various forms in recent work on multi-level synthesis 
\cite{mishchenko02boolean,mishchenko02encoding} using non-SOP encodings ranging from And-Inverter Gates (AIGs) and XOR-AND nets to graph-based representations in the tradition of \cite{bryant86graphbased}.
Interestingly, new synthesis targets, ranging from AIGs to cyclic combinational circuits \cite{riedel04cyclic}, turned out to be competitive with more traditional minimization based synthesis techniques. Synthesis of reversible circuits with possible uses in low-power adiabatic computing and quantum computing \cite{shende03synthesis} have emerged. Despite its super-exponential complexity, exact circuit synthesis efforts have been reported successful for increasingly large circuits \cite{drechsler98exact,knuth06draft}.
While \cite{reintjes92} describes the basics of CMOS technology, we refer the reader interested in full background information for our transistor models to \cite{rabaey}.

\section{Future Work} \label{fwork}
While we have provided unusually low cost transistor models for $<$ gates, the validation of their use in various context requires more extensive SPICE simulations as well precise area, delay and power estimates. 

The relative simplicity of $A<B$ suggests its use in novel analog or non-silicon designs provided that one can measure that signal A is in a given sense weaker than B. Its relative expressiveness challenges, to some extent, the widely believed statement \cite{dietmeyer71,symbool05} that symmetric functions are genuinly more interesting for circuit synthesis. Future work is needed to substantiate our belief that this is not necessarily the case, based on the intuition that asymmetric operators cover a larger combinatorial space than their associative/commutative siblings.

The {\em dual} of the $(<,1)$ library, $(\Rightarrow,0)$ has the same expressive power as $(<,1)$. It would be interesting to see if one can find similar low-transistor count implementations as the ones shown in this paper for $(<,1)$. Given that $(\Rightarrow,0)$ has been used as a foundation of various {\em implicative} formalizations of classic and intuitionistic logics, stronger rewriting mechanisms might be available for it than the ones we described in subsection \ref{rew} for $(<,1)$, resulting in better heuristics for handling circuits for which exact synthesis is intractable.

On the general synthesis algorithm side, it would be interesting to add tabling of subcircuits (through the use of a system like XSB or by writing a special purpose circuit store) to avoid recomputation. Adapting intelligent backtracking mechanisms like those used in modern SAT-solvers should also be considered to improve performance.

Given our focus -- to point out the usefulness of a relatively simple and unexplored primitive $<$ as a universal boolean function with small transistor count implementation -- we have not invested an implementation effort comparable to high quality synthesis tools like \cite{abc}. An interesting development would be adapting a tool like {\bf abc} \cite{abc} to specifically use $<$ as a primitive.

\section{Conclusion} \label{conc}
We have described a general logic programming based exact circuit synthesis algorithm and shown how Prolog language features like logic variables and backtracking can be used to provide efficient, a concise and elegant implementations. The synthesis algorithm has been used to identify the universal boolean function pair $(<,1)$ as a primitive for circuit synthesis, after noticing the possibility of a 4-transistor PTL-implementation. We have also shown that, surprisingly, despite being non-commutative and non-associative, Strict Boolean Inequality ``$<$" allows very low transistor-count implementations of typical small circuits. We have also provided a rewriting-based simplification algortithm in terms of $(<,1)$ that handles symbolic boolean expressions as well as CNF or DNF forms. This rewriting algorithm is usable for heuristic circuit synthesis when formula complexity makes exact synthesis intractable. We hope that these results will provide practical opportunities for the use of logic programming languages and their constraint handling extensions as components of circuit design automation software.

\bibliographystyle{plain}
\bibliography{syn}
 
\end{document}